\documentclass[english,aps,a4paper,twocolumn,showpacs,amsfonts,amsmath,amssymb,amscd,nofootinbib,superscriptaddress]{revtex4-1}
\usepackage[english]{babel}
\usepackage{graphicx}
\usepackage{hyperref}
\usepackage{amsmath}
\usepackage{amssymb}
\usepackage{enumerate}

\newcommand{\EE}{\mathbb{E}}

\newcommand{\PP}{\mathbb{P}}
\newcommand{\RR}{\mathbb{R}}
\newcommand{\TT}{\mathbb{T}}

\newcommand{\lint}{\ell}

\begin{document}
\def\Journal#1#2#3#4#5#6#7{{#1}, {#4} {\textbf{#5}}, #6 (#2)}
\def\Book#1#2#3#4#5{{#1}, in {\it #3} (#4, #5, #2).}

\title{Continuous-time random walks with reset events: Historical background and new perspectives}
\author{Miquel Montero}
\email[Corresponding author: ]{miquel.montero@ub.edu}
\affiliation{Departament de F\'{\i}sica de la Mat\`eria Condensada, Universitat de Barcelona (UB), Mart\'{\i} i Franqu\`es 1, E-08028 Barcelona, Spain}
\affiliation{Universitat de Barcelona Institute of Complex Systems (UBICS), Universitat de Barcelona, Barcelona, Spain}
\author{Axel Mas\'o-Puigdellosas}
\affiliation{Departament de F\'{\i}sica de la Mat\`eria Condensada, Universitat de Barcelona (UB), Mart\'{\i} i Franqu\`es 1, E-08028 Barcelona, Spain}
\author{Javier Villarroel}
\affiliation{Departamento de Matem\'aticas \& Instituto Universitario de F\'{\i}sica Fundamental y Matem\'aticas, Universidad de Salamanca, Plaza Merced s/n, E-37008 Salamanca, Spain}
\date{\today}
\begin{abstract}
In this paper, we consider a stochastic process that may experience random reset events which relocate the system to its starting position. We focus our attention on a one-dimensional, monotonic continuous-time random walk with a constant drift: the process moves in a fixed direction between the reset events, either by the effect of the random jumps, or by the action of a deterministic bias. However, the orientation of its motion is randomly determined after each restart.  As a result of these alternating dynamics, interesting properties do emerge. General formulas for the propagator as well as for two extreme statistics, the survival probability and the mean first-passage time, are also derived. The rigor of these analytical results is verified by numerical estimations, for particular but illuminating examples.
\end{abstract}
\pacs{02.50.Ey, 02.50.Ga, 05.40.Fb, 89.20.-a}%

\maketitle
\section{Introduction}
\label{sec:intro}

In 1965, 138 years after Robert Brown observed the random motion of a particle suspended on a fluid, in 1827, and 60 years after the Einstein description of its dynamics, in 1905, Montroll and Weiss considered the continuous-time random walk (CTRW) as a generalization of the original random walk where the time lapse between consecutive jumps is also a random variable with a given probability density. This was the starting point  to multitude developments on the physics of anomalous self-diffusion and applications to anomalous relaxation with power-law distributions.  Since then, several applications have been developed as in finance \cite{Sca07}, ecology~\cite{Vic14} or biology~\cite{Hof13}. Furthermore, new mechanisms have been included to the CTRW as, for instance, resets consisting in instantaneous relocations of the random walker to a given position.
    
But resets have also become part of  our daily lives: our search strategies when something is lost, the resetting of the router when the Internet signal is weak or the resetting of our computer when it is jammed. Even the periodicity of our sedentary lifestyle, returning to a fixed place at the end of the day to rest, can be interpreted as a reset. Crucially, in most of these situations, resets \textit{optimize} human activity: sometimes it is a time-saving strategy to restart an ineffective task from the beginning, and assume the delay associated with this decision, rather  than trying to fix it. This optimality and the existence of resets in natural processes are the main reasons that explain why including resets to certain processes has been a usual practice in different fields ranging from ecology to computer science.
    
At the end of the 1970s, stochastic resets were first studied in a mathematical sense within a continuous-time, discrete Markov process context as a tool to study the age of the processes by Levikson~\cite{Lev77}  and Pakes~\cite{Pak78,Pak79}. In these seminal papers, the authors concatenated multiple generations of a Markov chain to study the running-time distribution of a single process which indeed is equivalent to study a single process subjected to resets. Almost two decades later, in 1994, Kyriakidis \cite{Kir94} studied a more concrete application of these techniques in a population birth-death process with immigration and stochastic total catastrophes. Since then, a variety of different discrete Markov processes with resets have been proposed to model systems as, e.g., populations \cite{Swi01,Cha03} or queues \cite{Kri00,DiC03} |both in the discrete and  continuous limit. %on which they also study its continuous limit.
    
In the early 1990s great interest was raised to use  resets as a tool for optimizing algorithms. In their 1993 work, Luby, Sinclair and Zuckerman \cite{LSZ93} studied the efficiency of  Las Vegas search algorithms with restarts, and they showed the existence of an optimal resetting strategy. Their idea was based in the fact that the Las Vegas algorithm, due to its random nature, could become lost in some regions of the configuration space far from the actual solution, while a reset could help the algorithm recover the right path. Ulterior general studies in this field can be found in \cite{Kau02,Hua06}.
   
New seeds were sown among physicists in 1999 with the work of  Manrubia and Zanette \cite{Man99}, who found that when random resets are applied to a stochastic multiplicative process (SMP), power-law distributions do genuinely appear. More recently, in 2011, Evans and Majumdar \cite{Eva11_1,Eva11_2} studied a diffusing model with a resetting term in a Fokker-Planck equation, derived from microscopical considerations.  Afterwards, several analysis and generalizations of this formulation have been performed, including: The incorporation of an absorbing state \cite{Eva13_2}; the generalizations to $d$-spatial dimensions \cite{Eva14_1}; the presence of a general potential \cite{Pal14}; the inclusion of time dependency in the resetting rate \cite{Pal15} or a general distribution for the reset time \cite{Eul15}; a study of large deviations in Markovian processes \cite{Tou15}; a comparison with deterministic resetting \cite{Bha16}; the relocation to a previously position \cite{Boy17}; analyses on general properties of the first-passage time \cite{Reu16,Pal17}; or the possibility that internal properties drive the reset mechanism of the system \cite{Fal16}. 
   
Several discrete models have been analyzed under the presence of resets as in \cite{Jan12} where a random walker is forced to go back to the initial point when the mean hitting time of the process (the mean lapse to reach a fixed target, starting from the present location) is larger than the mean hitting time starting from the origin. Furthermore, the possibility of having walkers relocated to a known position \cite{Boy14} or, more specifically, to a previous maximum \cite{Maj15}, have been considered. Here we can also find the Sisyphus random walk \cite{Miq16}, where a walker with oriented and deterministic step lengths is subject to random resets to the initial point.
       
Resets have also been studied in more concrete applications as in L\'evy flights \cite{Kus14,Kus15,Vic15}, in coagulation-diffusion processes \cite{Dur13} or in the modeling of RNA polymerases to describe cleavages during the so-called backtracking, where the RNA performs a random walk to scan the DNA template \cite{Rol16}. Also, the thermodynamic properties of resetting stochastic processes have been widely studied in \cite{Fuc16}, while in \cite{Rot15%,Hus16
} general properties of restarted processes are analyzed by drawing an analogy with Michaelis-Menten reactions. %scheme. 
    
There have also been some recent works in which resets are superposed to the fundamental law that governs the CTRW. In \cite{Miq13}, an integral equation for the propagator of the system is found and solved in a direct way by taking into consideration some properties of the studied processes. Concretely, it is assumed that the walker motion follows from the conjunction of three different effects: a deterministic drift, instantaneous positive jumps, and reset events that bring the system to the origin. A different but related approach has been considered in \cite{Vic16} where some properties of the stationary state are derived for two different models: a \emph{jump} model where the walker jumps instantaneously from one state to another, and a \emph{velocity} model where the walker performs a ballistic movement.

The paper is organized as follows: In Sec.~\ref{sec:review} we introduce the general framework of the process under study, and review some of the results previously reported in~\cite{Miq13}. % a biased CTRW on which an exogenous restart mechanism operates, and review some previous results,. 
In Sec.~\ref{sec:alternate}  we generalize the dynamics of this model by allowing the system to change its direction after a reset, and obtain explicit analytic expressions for the transition probability of the system. Section~\ref{sec:extreme} is devoted to the analysis of the properties of two extreme-value statistics, the survival probability and the mean first-passage time.  We give a general formula for these statistics in terms of integral transforms, and some explicit examples are analyzed in detail. The paper ends with Sec.~\ref{sec:conclusions}, where conclusions are drawn.% where future perspectives are also sketched. %We have left for the appendices some technical mathematical derivations.

\section{CTRW with drift and resets}
\label{sec:review}

\begin{figure}[htbp]
\includegraphics[scale=0.65]{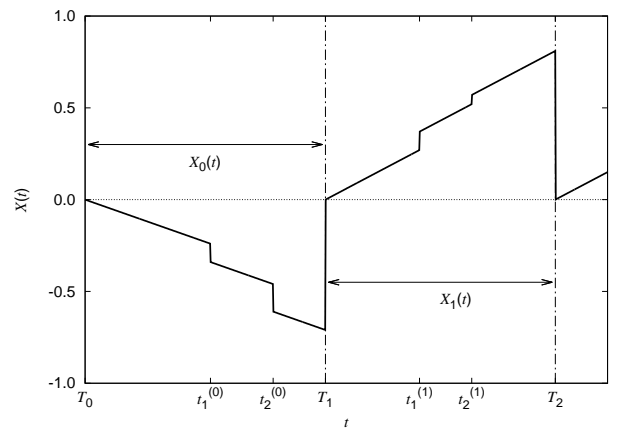}
\caption{Sample path of the processes $X(t)$. The solid line represents a possible realization of process $X(t)$ which grows linearly with velocity $\Gamma_m$ between the jump times $t^{(m)}_n$ and is instantaneously transferred to $X_0^{(m)}$ at the reset times $T_m$.} 
\label{fig:sample_path}
\end{figure}

Let us consider $X(t)$, a continuous-time random walk with drift which is susceptible of being instantaneously relocated. Let us denote by $\left\{T_m\right\}_{m=0,1,...}$ the set of %random 
reset times, and by
\begin{equation}
\left\{X_0^{(m)}\right\}_{m=0,1,...}\equiv\Big\{X(T_m)\Big\}_{m=0,1,...},
\end{equation}
the associted destiny locations. Under these premises, the process can be written as
\begin{equation}
X(t)=\sum_{m=0}^\infty X_m(t)\, \Theta(t-T_m)\left[1-\Theta(T_{m+1}-t)\right],
%\label{Z_def}
\end{equation}
with
\begin{eqnarray}
X_m(t)&\equiv& X_0^{(m)} + \Gamma_m \cdot (t-T_m)\nonumber\\
&+& \sum_{n=1}^\infty J_n^{(m)}\Theta\left(t-t_n^{(m)}\right)\Theta\left(t_n^{(m)}-T_m\right),
\label{X_m_def}
\end{eqnarray}
where $\Theta(t)$ is the right-continuous Heaviside step function, $\Theta(t)=1$ if $t\geq 0$ and zero otherwise; $\Gamma_m$ is the inter-reset drift velocity; and $J_n^{(m)}$ and $t_n^{(m)}$ are the jump sizes and jumping times of the CTRW, respectively |see Fig.~\ref{fig:sample_path}. 

%The parameters of this model are, in principle, independent and identically distributed (i.i.d.) random variables that can be split into two groups: global parameters and local parameters. The global parameters (the drift, the reset time and the reset position) are deterministic when looking the inter-reset period while the local parameters (the jump sizes and the jumping times) are stochastic within the inter-reset period. The local parameters are ruled by the probability density functions, $h^{(m)}(\cdot)$ and $\psi^{(m)}(\cdot)$, respectively:
The parameters of this model are, in a general sense, independent and identically distributed (i.i.d.) random variables that can be split into two groups: global parameters and local parameters. The global parameters (the drift, the reset time and the reset position) are deterministic when looking at some inter-reset period, while they are i.i.d. random variables when the system is considered globally. The local parameters (the jump sizes and the jumping times) are i.i.d. random variables when observed locally within a single inter-reset period. 
Local parameters are ruled by the probability density functions $h^{(m)}(\cdot)$ and $\psi^{(m)}(\cdot)$, respectively:
\begin{eqnarray}
h^{(m)} (u)du &\equiv & \PP\{ u<J_n^{(m)} \leq u+du\}, \label{h_m_def} \\
\psi^{(m)}(\tau)d\tau & \equiv & \PP\{\tau< \tau_n^{(m)} \leq \tau +d\tau \}, \label{psi_m_def}
\end{eqnarray} 
%which can in principle be different for different $m$, 
which can in principle vary with $m$, i.e., after a reset these distributions could change, depending even on the values that the global parameters take in $X_m(t)$. In Eqs.~(\ref{h_m_def}) and~(\ref{psi_m_def}) we have denoted by $\PP\{\cdots\}$ the probability of the set $\{\cdots\}$, and defined the waiting times between consecutive jumps, 
%\begin{equation*}
$\tau_n^{(m)}\equiv t_n^{(m)}-t_{n-1}^{(m)}$,
%\end{equation*}
with $t_0^{(m)}=T_m$. Therefore, local parameters can eventually depart from the i.i.d. assumption when considered globally.
 
%The global parameters are i.i.d. random variables as well. 
As done with the jump times, it is convenient to define the inter-reset waiting times, $\mathcal{T}_m\equiv T_m-T_{m-1}$, $T_0=0$. Then, the PDFs for the reset positions, the drift values and these reset waiting times are, respectively:
\begin{eqnarray}
g(x)d x & \equiv & \PP\{x< X_0^{(m)} \leq x +dx \},\\
\sigma(v)dv & \equiv &\PP \{ v<\Gamma_m \leq v+dv\}, \\
\phi (\tau)d\tau &\equiv & \PP\{ \tau< \mathcal{T}_m\leq\tau+d\tau \}.
\end{eqnarray} 
%Obviuously, waiting times must be positive. 
The process $X(t)$ consists then on concatenated sequences, of random length given by $\phi(\tau)$, of the usual continuous-time random walk, starting from different positions given by $g(x)$, and with different drift velocities given by $\sigma(v)$. %(Fig. \ref{fig1}). 

Within the present framework, a particular case has been studied in Ref.~\cite{Miq13}, where only positive drifts and jumps are considered. To be precise, in Ref.~\cite{Miq13} the reset positions for the walker are chosen to be the origin, a common drift to all different inter-reset intervals, which are exponentially distributed, that is,
\begin{eqnarray}
g(x)&=&\delta(x),\\
\sigma(v)&=&\delta(v-\Gamma), \\
\phi(\tau)&=&\Lambda e^{-\Lambda\tau},\label{Poisson_L}
\end{eqnarray}
with $\Gamma>0$, and being $\Lambda^{-1}$ the mean inter-reset time. The local random variables are drawn in each inter-reset period from the same distributions, i.e., $h^{(m)}(u)= h(u)$, with $h(u)=0$ for $u<0$, and  $\psi^{(m)}(\tau)= \psi(\tau)$, 
\begin{equation}
\psi(\tau)=\lambda e^{-\lambda \tau},
\label{Poisson_l}
\end{equation}
the probability density corresponding to a Poisson point process, with intensity $\lambda$.

Under these assumptions, the process describes a directed motion which is suddenly reseted to the origin, at random times $T_m$. Note that, while in the general setup the system can have long memory, $X(t)$ becomes now time-homogeneous and Markovian.  The evolution of the process is completely determined by means of its propagator:
%The process can be studied by means of its propagator 
\begin{equation}
p(x,t;x_0,t_0)\equiv\PP \left\{ x<X(t)\leq x+dx|X(t_0)=x_0\right\}.
\end{equation}
This is not translationally invariant since the resets break this symmetry, i.e.,
\begin{equation}
p(x,t;x_0,t_0)=p(x,t-t_0;x_0,0)\equiv p(x,\tau;x_0),
\end{equation}
where $t_0$ and $t$ (with $t_0\leq t $) are, respectively, the \emph{present} time and a given future instant;  $\tau\equiv t-t_0$ is the associated time lapse.

We can now derive an integral equation taking into account the features of the model. Three are the different and mutually exclusive scenarios which we have to take into account depending on the interval time $\tau$, the first reset time $\tau'$ and the first jump time $\tau''$|note that $\tau'$ and $\tau''$ are random variables while $\tau$ is just a number: \textbf{(i)}  both the first jump time and the first reset time are larger than the interval time; \textbf{(ii)} the first reset happens within interval $\tau$ and before any jump; or \textbf{(iii)} the first jump happens within interval $\tau$ and before any reset. This three possible events lead to
\begin{eqnarray}
p(x,\tau;x_0)&=& e^{-(\lambda+ \Lambda)\tau}\delta(x-x_0-\Gamma \tau)  \nonumber \\
&+&\int_0^\tau d\tau' \Lambda e^{-(\lambda+\Lambda)\tau'}p(x,\tau-\tau';0)  \nonumber \\
&+&\int_0^\infty duh(u)\int_0^\tau d\tau'' \lambda e^{-(\lambda+\Lambda)\tau''}\nonumber \\
&\times&p(x,\tau-\tau'';x_0+\Gamma\tau''+u),
\label{eq3}
\end{eqnarray}
whose solution in the Laplace space, both for spatial and temporal variables, reads
\begin{equation}
\hat{\hat{p}}(r,s;x_0)= \frac{\frac{\Lambda}{s}+e^{-r x_0}}{s + \Lambda +\lambda[1-\hat{h}(r)]+\Gamma r},
\label{eq4}
\end{equation}
where 
\begin{eqnarray}
\hat{\hat{p}}(r,s;x_0)&\equiv&\int_0^\infty dx\,e^{-r x} \int_0^\infty d\tau\,e^{-s \tau} p(x,\tau;x_0),\\
\hat{h}(r)&\equiv&\int_0^\infty du\,h(u)\,e^{-r u}.
\end{eqnarray}
For an exhaustive derivation of the propagator structure and a wide discussion on the method used to solve Eq.~\eqref{eq3} we refer the reader to the original work \cite{Miq13}, where the ergodicity and the existence of a stationary density for the process are also proved. In particular,  for slow resetting rates, $\Lambda \ll \lambda$ and $\Lambda \ll \Gamma$, the tail of the stationary density for the exponentiated process $Y(t)$,  $Y(t)\equiv e^{X(t)}$, is found to decay as
\begin{equation}
p_Y(y)\propto \frac{\beta}{y^{1+\beta}},
\end{equation}
with $\beta \equiv \Lambda/(\Gamma +\lambda\, \EE[J])$,
%\begin{equation}
%\beta \equiv \frac{\Lambda}{ \Gamma +\lambda\, \EE[J]},
%\end{equation}
which depends explicitly on $\Lambda$, $\lambda$, $\Gamma$, and the first moment of the jump distribution $h(u)$,
\begin{equation}
 \EE[J]\equiv\int_0^\infty u\, h(u) du.
\end{equation}
This result corroborates the interesting finding of  Ref.~\cite{Man99} to the extent that  the inclusion of resets in SMP leads naturally to power-law distributions.

\section{Alternate process with resets}
\label{sec:alternate}
The alternate model described in this section is a rich generalization of the previous case and consists in a sequence of monotonic processes which are decided to be positively or negatively oriented with a given probability, $\rho$ and $1-\rho$, respectively. This type of process may lead to an optimal resetting rate for the first-passage time as shown for the discrete model in \cite{Miq16}, since now a reset does not always represent a penalty for the process: it can help the walker to regain its direction when it has reached a region which is far from its target. 

As in the monotonic model, the waiting times are exponentially distributed, and we fix the destination of the reset mechanism to be always the origin, $g(u)=\delta(u)$. By contrast, for the probability density of the drift we have
\begin{equation}
\sigma(v)=\rho\, \delta(v-\Gamma_+)+(1-\rho)\delta(v+\Gamma_-),
\label{eq20}
\end{equation}
and the jump-size distribution must be chosen accordingly, i.e., $h^{(m)}(u)=h_+(u)$ if $\Gamma_{m}=\Gamma_+$, and $h^{(m)}(u)=h_-(u)$ if $\Gamma_{m}=-\Gamma_-$. Note that, for the sake of readability, we have implicitly denoted by $\Gamma_{\pm}$ the \emph{modulus} of the velocity, $\Gamma_{\pm}\geq 0$, that is, the system constant speed. Thus, the process will drift into either the positive or the negative region, depending on the contextual subscript ($\pm$). 

We will proceed in the same way with the jump magnitudes, by requiring that $J^{(m)}_n\geq 0$, or, in other words that, $h_{\pm}(u)=0$ for $u<0$. This introduces a small modification in the formal definition of $X_m(t)$ in Eq.~\eqref{X_m_def}, the expression that determines the evolution of the process during the inter-reset intervals, which here reads
%\begin{eqnarray}
%X_{m}(t)&=&\pm \Gamma_{\pm} \cdot (t-T_m)\nonumber\\
%&\pm& \sum_{n=1}^\infty J_n^{(m)}\Theta\left(t-t_n^{(m)}\right)\Theta\left(t_n^{(m)}-T_m\right).
%\end{eqnarray}
\begin{equation}
X_{m}(t)=\pm \left[\Gamma_{\pm} \cdot (t-T_m) + \sum_{n=1}^\infty J_n^{(m)}\Theta\left(t-t_n^{(m)}\right)\right].
\end{equation}
This is the case depicted in Fig.~\ref{fig:sample_path}.

We are now in a position to analyze the probabilistic properties of this alternating process. Let us derive first the integral equation that governs the evolution of $p(x,\tau;x_0)$. To this end, we will introduce the conditional functions $p_{\pm}(x,\tau;x_0)$,
%\begin{equation}
%p_{\pm}(x,\tau;x_0)\equiv p(x,\tau;x_0)\Theta(\pm x_0),
%\end{equation}
two functions that depend on the present direction of the movement of the system. The reason behind this apparently redundant definition (the system is moving rightwards if $x_0>0$ and leftwards if $x_0<0$) can be traced to the singular nature of the origin  $x_0=0$, at which one has (in general) that $p_{+}(x,\tau;0)\neq p_{-}(x,\tau;0)$.
%The reason behind this apparently redundant definition (the system is moving rightwards if $x_0>0$ and leftwards if $x_0<0$) is hidden in the singular case of $x_0=0$, for which one has, in general, that $p_{+}(x,\tau;0)\neq p_{-}(x,\tau;0)$.

 As before, we denote by $\tau'$ and $\tau''$ the time interval up to the first reset event and, respectively, the first jump. The equation for $p_{\pm}(x,\tau;x_0)$ can be built up by considering the three possible and mutually exclusive scenarios that appear depending on the relative values of the three intervals $\tau$, $\tau'$, and $\tau''$:
\begin{enumerate}[\bf (i)]
\item %(i) 
There is neither a reset nor a jump in the time interval $\tau$, i.e., $\tau'>\tau$ and $\tau''>\tau$. In this case, depending on the system \textit{inertia}, one has $X(\tau)=x_0\pm \Gamma_{\pm} \tau$.  System may reach the point $x$ at this stage, only if $x \cdot x_0 \geq 0$. 
\item %(ii) 
There is at least one reset in the time interval, and the first one takes place before any jump has occurred, $\tau'\leq\tau$ and $\tau''>\tau'$.  In this case the transition PDF after the reset will be $p_{+}(x,\tau-\tau';0)$, with probability $\rho$, or $p_{-}(x,\tau-\tau';0)$, with probability $(1-\rho)$. %Starting from the origin, the system must reach $x$, in a lapse of time $\tau-\tau'$.
\item %(iii) 
There is at least one jump in the considered interval, $\tau''\leq\tau$. This first jump takes place before the first reset event takes place, $\tau'>\tau''$. Right after the jump we have $X(\tau'')=x_0\pm\Gamma_{\pm} \tau'' \pm u$, where the magnitude $u$ of the jump is drawn from the density $h_{\pm}(\cdot)$, depending on the direction of the movement.
The propagator is then $p(x,\tau-\tau'';x_0\pm \Gamma_{\pm} \tau''\pm u)$.
\end{enumerate}
In view of all this $p_{\pm}(x,\tau;x_0)$ must satisfy the following renewal equation:

\begin{widetext}
\begin{eqnarray}
p_{\pm}(x,\tau;x_0)&=&\int_\tau^{\infty}d\tau'\Lambda e^{-\Lambda \tau'}\int_\tau^{\infty}d\tau'' \lambda_{\pm} e^{-\lambda_{\pm} \tau''} \delta(x-x_0\mp \Gamma_\pm \tau) \Theta(\pm x_0) \nonumber \\
&+&\int_0^{\tau}d\tau'\Lambda e^{-\Lambda \tau'}\int_{\tau'}^{\infty}d\tau'' \lambda_{\pm}  e^{-\lambda_{\pm}  \tau''} \Big[\rho\, p_+(x,\tau-\tau';0) + (1-\rho)\, p_-(x,\tau-\tau';0)\Big]
\nonumber\\
&+&\int_0^{\tau}d\tau''\lambda_{\pm}  e^{-\lambda_{\pm} \tau''}\int_{\tau''}^{\infty}d\tau' \Lambda e^{-\Lambda \tau'} \int_{0}^{\infty}du h_{\pm} (u)p_{\pm} (x,\tau-\tau'';x_0\pm \Gamma_{\pm}  \tau''\pm u)\nonumber\\
&=& e^{-(\Lambda+\lambda_{\pm}) \tau} \delta(x-x_0\mp\Gamma_{\pm}  \tau)\Theta({\pm} x_0) +\int_0^{\tau}d\tau'\Lambda e^{-(\Lambda+\lambda_{\pm}) \tau'}  \Big[\rho\, p_+(x,\tau-\tau';0) + (1-\rho)\, p_-(x,\tau-\tau';0)\Big]
\nonumber\\
&+&\int_{0}^{\infty}du h_{\pm} (u) \int_0^{\tau}d\tau''\lambda e^{-(\Lambda+\lambda_{\pm}) \tau''} p(x,\tau-\tau'';x_0\pm\Gamma_{\pm}  \tau''\pm u).%\nonumber\\
\label{TPDF}
\end{eqnarray}
\end{widetext}

As stated in the previous section, the standard procedure for solving integral equations like~(\ref{TPDF}) is to resort to the use of some integral transformation, either the Laplace transform, the Fourier transform, or a combination of them. Here, since $x\in \RR$ and $\tau\geq0$, the natural choice is to consider the Fourier transform in the first argument and the Laplace transform in the second one:
\begin{equation}
\hat{\tilde{p}}_{\pm}(\omega,s;x_0)\equiv\int_0^{\infty}d\tau e^{-s \tau} \int_{-\infty}^{\infty} dx\, p_{\pm}(x,\tau;x_0) e^{i \omega x},
\end{equation}
where %from now on 
the hat will denote the Laplace transform with respect to the {\it time\/} variable, and the tilde will denote the Fourier transform with respect to the {\it position\/} variable. In the most typical situation, the system of equations obtained after such integral transformation can be solved through direct algebraic manipulation. In the present case, however, the problem in the Fourier-Laplace space is simpler, but $\hat{\tilde{p}}_{\pm}(\omega,s;x_0)$ is still the solution of an integral equation: 

%\begin{widetext}
%\begin{eqnarray}
%\hat{\tilde{p}}_{\pm}(\omega,s;x_0)
%&=&\frac{1}{s +\Lambda+\lambda_{\pm}\mp i\omega \Gamma_{\pm}}e^{i\omega x_0} \Theta(\pm x_0)%\nonumber\\&+&
%+\int_0^{\infty}d\tau''\lambda e^{-(s+\Lambda+\lambda_{\pm}) \tau''} %\nonumber\\ &\times &
%\int_{0}^{\infty}du h_{\pm}(u)\hat{\tilde{p}}(\omega,s;x_0\pm\Gamma_{\pm} \tau''\pm u) \nonumber\\ 
%&+&\frac{\Lambda}{s + \Lambda +\lambda_{\pm}}\Big[\rho\,\hat{\tilde{p}}_+(\omega,s;0)+ (1-\rho)\, \hat{\tilde{p}}_-(\omega,s;0)\Big].
%\nonumber\\
%\label{LTPDF}
%\end{eqnarray}
%\end{widetext}

\begin{eqnarray}
\hat{\tilde{p}}_{\pm}(\omega,s;x_0)
&=&\frac{1}{s +\Lambda+\lambda_{\pm}\mp i\omega \Gamma_{\pm}}e^{i\omega x_0} \Theta(\pm x_0)\nonumber\\
&+&\frac{\Lambda\Big[\rho\,\hat{\tilde{p}}_+(\omega,s;0)+ (1-\rho)\, \hat{\tilde{p}}_-(\omega,s;0)\Big]}{s + \Lambda +\lambda_{\pm}}\nonumber\\ 
&+&\int_0^{\infty}d\tau''\lambda e^{-(s+\Lambda+\lambda_{\pm}) \tau''} \nonumber\\ 
&\times &\int_{0}^{\infty}du h_{\pm}(u)\hat{\tilde{p}}(\omega,s;x_0\pm\Gamma_{\pm} \tau''\pm u). \nonumber\\
\label{LTPDF}
\end{eqnarray}

We faced a similar problem in Ref.~\cite{Miq13}. Based on our previous expertise, we posit the ansatz
\begin{eqnarray}
\hat{\tilde{p}}_{\pm}(\omega,s;x_0)&=&\hat{\tilde{q}}_{\pm}(\omega,s\mp i\omega \Gamma_{\pm})e^{i\omega x_0} \Theta(\pm x_0) \nonumber \\
&+&\frac{\Lambda \rho}{s} \hat{\tilde{q}}_{+}(\omega,s- i\omega \Gamma_+) \nonumber \\
&+&\frac{\Lambda (1-\rho)}{s} \hat{\tilde{q}}_{-}(\omega,s+ i\omega \Gamma_{-}),
\label{Ansatz_p}
\end{eqnarray}
where $\hat{\tilde{q}}_{\pm}(\omega,s)$ are two auxiliary functions to be determined. By insertion of (\ref{Ansatz_p}) into the right-hand side of Eq.~(\ref{LTPDF}) we obtain
\begin{eqnarray}
\hat{\tilde{p}}_{\pm}(\omega,s;x_0)
&=&\frac{e^{i \omega x_0} \Theta(\pm x_0)}{s+\Lambda+\lambda_{\pm }\mp i \omega\Gamma_{\pm}}\nonumber\\
&+&\frac{\lambda_{\pm}e^{i \omega x_0} \Theta(\pm x_0) }{s+\Lambda+\lambda_{\pm }\mp i \omega\Gamma_{\pm}}\tilde{h}_{\pm}(\omega)\hat{\tilde{q}}_{\pm}(\omega,s\mp i\omega \Gamma_{\pm}) \nonumber\\
&+&\frac{\Lambda \rho}{s} \hat{\tilde{q}}_{+}(\omega,s- i\omega \Gamma_+) \nonumber \\
&+&\frac{\Lambda (1-\rho)}{s} \hat{\tilde{q}}_{-}(\omega,s+ i\omega \Gamma_{-}),
\label{Ansatz_LTPDF}
\end{eqnarray}
which is self-consistent with Eq.~(\ref{Ansatz_p}) if and only if
\begin{equation}
\hat{\tilde{q}}_{\pm}(\omega,s)=\frac{1}{s+\Lambda+\lambda_{\pm }\left[1-\tilde{h}_{\pm}(\omega)\right]},
\label{Ansatz_sol}
\end{equation}
where we have defined
\begin{equation}
\tilde{h}_{\pm}(\omega)\equiv\int_0^{\infty}du\, h_{\pm}(u) e^{\pm i \omega u}.
\end{equation}
We can finally substitute formula~(\ref{Ansatz_sol}) in Eq.~(\ref{Ansatz_p}) in order to get the explicit expression of $\hat{\tilde{p}}_{\pm}(\omega,s;x_0)$,
\begin{eqnarray}
\hat{\tilde{p}}_{\pm}(\omega,s;x_0)&=&\frac{e^{i \omega x_0} \Theta(\pm x_0)}{s+\Lambda+\lambda_{\pm }\left[1-\tilde{h}_{\pm}(\omega)\right]\mp i \omega\Gamma_{\pm}}\nonumber \\
&+&\frac{\Lambda \rho}{s} \frac{1}{s+\Lambda+\lambda_{+}\left[1-\tilde{h}_{+}(\omega)\right]- i \omega\Gamma_{+}}\nonumber \\
&+&\frac{\Lambda (1-\rho)}{s} \frac{1}{s+\Lambda+\lambda_{-}\left[1-\tilde{h}_{-}(\omega)\right]+ i \omega\Gamma_{-}}.\nonumber \\
\label{p_sol}
\end{eqnarray}
The propagator $p_{\pm}(x,\tau;x_0)$ then follows by Fourier-Laplace  inversion of this expression.

The recurrent nature of the dynamics governing our system leads to the existence of  a stationary PDF, which can be recovered from Eq.~\eqref{p_sol},
\begin{eqnarray}
\tilde{p}(\omega)&=&\lim_{\tau\rightarrow \infty}\tilde{p}_{\pm}(\omega,\tau;x_0)=\lim_{s\rightarrow 0}s\, \hat{\tilde{p}}_{\pm}(\omega,s;x_0)\nonumber\\
&=& \frac{\Lambda \rho}{\Lambda+\lambda_{+}\left[1-\tilde{h}_{+}(\omega)\right]- i \omega\Gamma_{+}}\nonumber \\
&+&\frac{\Lambda (1-\rho)}{\Lambda+\lambda_{-}\left[1-\tilde{h}_{-}(\omega)\right]+ i \omega\Gamma_{-}}.%\nonumber \\
\label{p_st}
\end{eqnarray}
Let us exemplify the Fourier inversion of this expression by assuming that the movement of the system is driven by exponentially-distributed jumps in the positive axis, and by a continuous drift in the negative axis, that is $\Gamma_+=0$, $\lambda_+=\lambda$, $h_+(u)=\gamma e^{-\gamma u}$, $\lambda_-=0$, and $\Gamma_-=\Gamma$. Under these assumptions the stationary distribution is a Laplace (or doubly exponential) distribution with %skewness, and 
a point mass at the origin:
\begin{eqnarray}
p(x)&=&\frac{\Lambda \rho}{\Lambda+\lambda}\left[\delta(x)+\frac{\gamma\lambda}{\Lambda+\lambda}e^{-\frac{\Lambda\gamma x}{\Lambda+\lambda}}\Theta(x)\right]\nonumber\\
&+&\frac{\Lambda (1-\rho)}{\Gamma}e^{\frac{\Lambda x}{\Gamma}}\Theta(-x).
\label{p_h_exp}
\end{eqnarray}

\section{Extreme-event statistics}
\label{sec:extreme}

In this Section we analyze some statistical properties associated to an \emph{extreme} event of the process $X(\tau)$: specifically, the first time that the process crosses a given level. %$\ell$. 
%First passage time of the system throught $\ell$, $\ell>0$...

To tackle this problem, let us introduce  the survival probability (SP) of the process, $\mathcal{P}_{\lint}(\tau;x_0)$,  
\begin{equation}
\mathcal{P}_{\lint}(\tau;x_0)\equiv \PP\left\{\left. X(\bar{\tau}) \leq \ell, \bar{\tau} \leq \tau \right| X(0)=x_0\right\},
\end{equation}
namely, the probability that the process, which is initially in $x_0$, $ x_0\leq \ell$, $\ell\geq 0$, does not leave the interval $(-\infty,\ell]$ before time $\tau$. Then, if we denote by $\mathbf{T}_{\lint}$ the first time the process traverses the threshold $\ell$, i.e., the first time the process exits $(-\infty,\ell]$, starting from $x_0$, 
\begin{equation}
\mathbf{T}_{\lint}\equiv\min \left\{\tau: X\left(\tau\right)\notin(\infty,\ell]| X(0)=x_0 \right\},
\label{Exit_time_def}
\end{equation}
then the SP %survival probability    
is simply the probability that $\mathbf{T}_{\lint}>\tau$, that is
\begin{equation}
\PP\left\{\mathbf{T}_{\lint}\leq\tau\right\}=1-\mathcal{P}_{\lint}(\tau;x_0).
\label{survival}
\end{equation}
A related magnitude of interest is the mean first-passage time (MFPT), the expected value of $\mathbf{T}_{\lint}$:
\begin{equation}
\TT_{\lint}(x_0)\equiv \EE\left[\mathbf{T}_{\lint}\right]=\int_0^{\infty} \mathcal{P}_{\lint}(\tau;x_0)\,d\tau.
\end{equation}

Note that, as long as the process alternates monotonic behavior, knowing the precise dynamics of the system for $x<0$, when $\ell>0$, is not necessary: the system can only reach the boundary while moving rightwards.~\footnote{Obviously, if we want to analyze the first passage throughout $\ell$, with $\ell<0$, the interval to be considered is $[\ell,+\infty)$, and the results will become independent of the dynamics of the system for $x>0$.}  Therefore, to ease the notation, in the sequel we drop the subscript in $\Gamma_+$ and $h_{+}(u)$, as well as we denote by simply $x$ the initial location of the system. In order to get an equation for $\mathcal{P}_{\lint}(\tau;x)$ we will resort to renewal arguments, very similar to those enumerated in Sec.~\ref{sec:alternate}, but with the additional requirement that the process does not cross the  boundary at any time. One more time, it is necessary to distinguish between the two cases: when the system is moving rightwards, $\mathcal{P}_{\lint}(\tau;x,+)$, and when the system is moving leftwards $\mathcal{P}_{\lint}(\tau;x,+)$. The reason again is that $\mathcal{P}_{\lint}(\tau;0,+)\neq\mathcal{P}_{\lint}(\tau;0,-)$.

With the same notation as in Sec.~\ref{sec:alternate}, we will explore the three non-overlapping situations for $\mathcal{P}_{\lint}(\tau;x,+)$, $x\geq0$:
\begin{enumerate}[\bf (i)]
\item There is neither a jump nor a reset in the time interval $\tau$, $\tau'>\tau$ and $\tau''>\tau$, and the elapsed time is not long enough to reach the boundary by the effect of the drift, $\tau\leq (\ell-x)/\Gamma$. In this case the process will survive with certainty.  
\item There is at least one reset event at $\tau'$ in the time interval, $\tau' \leq \tau$, before the first jump $\tau'<\tau''$. If $\tau'> (\ell-x)/\Gamma$ the process does not survive, otherwise the SP is $\mathcal{P}_{\lint}(\tau-\tau';0,+)$, with probability $\rho$, and $\mathcal{P}_{\lint}(\tau-\tau';0,-)$, with probability $(1-\rho)$.
\item There is at least one jump at $\tau''$, $\tau''<\tau'$, before the first reset, $\tau'' \leq \tau$. After this event there is survival probability $\mathcal{P}_{\lint}(\tau-\tau'';x+\Gamma \tau'+u)$ if and only if the process has not crossed the level either balistically, $\tau''\leq (\ell-x)/\Gamma$, or as a consequence of the jump, i.e., the jump size $u$ must be smaller than $\ell-x-\Gamma \tau''$.
\end{enumerate}
The case of  $\mathcal{P}_{\lint}(\tau;x,-)$, $x\leq 0$, is simpler: 
\begin{enumerate}[\bf (i)]
\item If there is no reset in the time interval $\tau$, $\tau'>\tau$, the process will survive with certainty.  
\item If there is at least one reset event at $\tau'$, $\tau'\leq\tau$,  the SP is $\mathcal{P}_{\lint}(\tau-\tau';0,+)$, with probability $\rho$, and $\mathcal{P}_{\lint}(\tau-\tau';0,-)$, with probability $(1-\rho)$.
\end{enumerate}
The above contingencies lead to the following set of coupled integral equations:

\begin{widetext}
\begin{eqnarray}
\mathcal{P}_{\lint}(\tau;x,+)&=&\int_\tau^{\infty}d\tau'\Lambda e^{-\Lambda \tau'} \int_\tau^{\infty}d\tau'' \lambda e^{-\lambda \tau''}\Theta(\ell-x-\Gamma \tau)
\nonumber\\
&+&\int_0^{\tau}d\tau'\Lambda e^{-\Lambda \tau'}\int_{\tau'}^{\infty}d\tau'' \lambda e^{-\lambda \tau''} \left[\rho\,\mathcal{P}_{\lint}(\tau-\tau';0,+)+(1-\rho)\,\mathcal{P}_{\lint}(\tau-\tau';0,-)\right]\Theta(\ell-x-\Gamma \tau')
\nonumber\\
&+&\int_0^{\tau}d\tau''\lambda e^{-\lambda \tau''}\int_{\tau''}^{\infty}d\tau' \Lambda e^{-\Lambda \tau'} \int_{0}^{\ell-x-\Gamma \tau''}du h(u)\mathcal{P}_{\lint}(\tau-\tau'';x+\Gamma \tau''+u,+) \Theta(\ell-x-\Gamma \tau'')
\nonumber\\
&=& e^{-(\Lambda+\lambda) \tau} \Theta(\ell-x-\Gamma \tau)%\nonumber\\&+&
+\int_0^{\tau}d\tau'\Lambda e^{-(\Lambda+\lambda) \tau'} \left[\rho\,\mathcal{P}_{\lint}(\tau-\tau';0,+)+(1-\rho)\,\mathcal{P}_{\lint}(\tau-\tau';0,-)\right]\Theta(\ell-x-\Gamma \tau')\nonumber\\
&+&\int_0^{\tau}d\tau''\lambda e^{-(\Lambda+\lambda) \tau''} \int_{0}^{\ell-x-\Gamma \tau''}du h(u)\mathcal{P}_{\lint}(\tau-\tau'';x+\Gamma \tau''+u,+) \Theta(\ell-x-\Gamma \tau'')
\\
\mathcal{P}_{\lint}(\tau;x,-)&=& e^{-\Lambda \tau} 
+\int_0^{\tau}d\tau'\Lambda e^{-\Lambda \tau'} \left[\rho\,\mathcal{P}_{\lint}(\tau-\tau';0,+)+(1-\rho)\,\mathcal{P}_{\lint}(\tau-\tau';0,-)\right],
\end{eqnarray}
\end{widetext}
\noindent
where, as it can be seen 
$\hat{\mathcal{P}}_{\lint}(\tau;x,-)$ does not depend on the precise value of 
$x$.

Next, we will consider the Laplace transform of the SP with respect to the time variable, 
\begin{equation}
\hat{\mathcal{P}}_{\lint}(s;x,\pm)\equiv\int_0^{\infty}d\tau \mathcal{P}_{\lint}(\tau;x,\pm) e^{-s \tau},
\label{LSP_def}
\end{equation}
and obtain
\begin{eqnarray}
\hat{\mathcal{P}}_{\lint}(s;x,+)&=&\frac{1- e^{-\frac{s+\Lambda+\lambda}{\Gamma}(\ell-x)}}{s+\Lambda+\lambda}\bigg\{1\nonumber\\
&+&\Lambda \left[\rho\, \hat{\mathcal{P}}_{\lint}(s;0,+)+(1-\rho)\, \hat{\mathcal{P}}_{\lint}(s;0,-)\right]\bigg\} 
\nonumber\\
&+&\frac{\lambda}{\Gamma}\int_0^{\ell-x}dz e^{-\frac{s+\Lambda+\lambda}{\Gamma}(\ell-x-z)} \nonumber\\
&\times&\int_{0}^{z}du h(u)\hat{\mathcal{P}}_{\lint}(s;\ell-z+u,+),
\\\label{Full_PLP}
\hat{\mathcal{P}}_{\lint}(s;x,-)&=&\frac{1}{s+\Lambda}\nonumber\\
&+&\frac{\Lambda}{s+\Lambda} \left[\rho\, \hat{\mathcal{P}}_{\lint}(s;0,+)+(1-\rho)\, \hat{\mathcal{P}}_{\lint}(s;0,-)\right].\nonumber\\
\label{Full_PLM}
\end{eqnarray}
Since $\hat{\mathcal{P}}_{\lint}(s;x,-)=\hat{\mathcal{P}}_{\lint}(s;0,-)$, from Eq.~(\ref{Full_PLM}) one obtains
\begin{eqnarray}
\hat{\mathcal{P}}_{\lint}(s;0,-)=\frac{1}{s+\Lambda\rho}\left[1+\Lambda\rho \hat{\mathcal{P}}_{\lint}(s;0,+)\right],
\end{eqnarray}
and thus
\begin{eqnarray}
\hat{\mathcal{P}}_{\lint}(s;x,+)&=&\frac{1- e^{-\frac{s+\Lambda+\lambda}{\Gamma}(\ell-x)}}{s+\Lambda+\lambda}\cdot\frac{s+\Lambda}{s+\Lambda\rho}\nonumber\\
&\times& \left[1+\Lambda\rho\, \hat{\mathcal{P}}_{\lint}(s;0,+)\right]
\nonumber\\
&+&\frac{\lambda}{\Gamma}\int_0^{\ell-x}dz e^{-\frac{s+\Lambda+\lambda}{\Gamma}(\ell-x-z)} \nonumber\\
&\times&\int_{0}^{z}du h(u)\hat{\mathcal{P}}_{\lint}(s;\ell-z+u,+).
\label{Full_PL}
\end{eqnarray}

Note that in this case we cannot perform any integral transform with respect to $x$ in a straightforward way because %$x\in [0,b]$
$x$ is restricted to the interval $[0,\ell]$. %In the following section we show how to overcome this limitation.
We also remind that by setting $s=0$ in Eq.~(\ref{Full_PL}) one obtains the corresponding integral equation for the mean exit time of the process out of the interval, $\TT_{\lint}(x,+)$:  %is directly obtained:%~\footnote{One can alternatively obtain this equation by means of the same kind of arguments that led to Eq.~(\ref{Full_PL}), see, e.g., \cite{mmp05,MV10}.}
\begin{eqnarray}
\TT_{\lint}(x,+)&=&\frac{1- e^{-\frac{\Lambda+\lambda}{\Gamma}(\ell-x)}}{\lambda+\Lambda}\left[\frac{1}{\rho}+\Lambda\, \TT_{\lint}(0,+) \right] 
\nonumber\\
&+&\frac{\lambda}{\Gamma}\int_0^{\ell-x}dz e^{-\frac{\Lambda+\lambda}{\Gamma}(\ell-x-z)} \nonumber \\
&\times&\int_{0}^{z}du h(u)\TT_{\lint}(\ell-z+u,+).
\label{Full_TL}
\end{eqnarray}

Here we consider the solution to Eq.~(\ref{Full_PL}) with arbitrary choice of the jump size PDF $h(\cdot)$. To this end we have to find the general solution of the allied integral equation 
\begin{eqnarray}
\hat{\mathcal{F}}(s;z)&=&\frac{\hat{\mathcal{A}}(s)}{s+\Lambda+\lambda}\left[1- e^{-\frac{s+\Lambda+\lambda}{\Gamma}z}\right] 
\nonumber\\
&+&\frac{\lambda}{\Gamma}\int_0^z dz' e^{-\frac{s+\Lambda+\lambda}{\Gamma}(z-z')}\nonumber\\
&\times&\int_{0}^{z'}du h(u)\hat{\mathcal{F}}(s;z'-u),
\label{Full_FL}
\end{eqnarray}
for $z\geq 0$, with $\hat{\mathcal{A}}(s)$ an arbitrary function of $s$. %~\footnote{Note that in Eq.~\eqref{Full_PL} $s$ is a parameter and hence to all effects $\hat{\mathcal{A}}(s)$ plays the role of a constant.} 
Notice that one recovers $\hat{\mathcal{P}}_{\lint}(s;x,+)$ via $\hat{\mathcal{P}}_{\lint}(s;x,+)=\hat{\mathcal{F}}(s;\ell-x)$ and requiring that 
\begin{equation}
\hat{\mathcal{A}}(s)=\frac{s+\Lambda}{s+\Lambda\rho}\left[1+ \Lambda\rho\, \hat{\mathcal{P}}_{\lint}(s;0,+)\right]. 
\end{equation}
The solution of $\hat{\mathcal{F}}(s;z)$ for values larger than $\ell$, i.e., when $x<0$, is mathematically meaningful but has no physical significance, since in this case the survival probability is not $\hat{\mathcal{P}}_{\lint}(s;x,+)$ but $\hat{\mathcal{P}}_{\lint}(s;x,-)$.

Expression~(\ref{Full_FL}) is now well suited to be Laplace transformed in the $z$ variable as well,
\begin{equation}
\hat{\hat{\mathcal{F}}}(s;r)\equiv \int_0^{\infty}dz \hat{\mathcal{F}}(s;z) e^{-r z},
\end{equation}
yielding in this way
\begin{eqnarray}
\hat{\hat{\mathcal{F}}}(s;r)&=&\frac{1}{s+\Lambda+\lambda+\Gamma r}\cdot\frac{\hat{\mathcal{A}}(s)}{r}\nonumber\\
&+&\frac{\lambda}{s+\Lambda+\lambda+\Gamma r} \hat{h}(r) \hat{\hat{\mathcal{F}}}(s;r).
\end{eqnarray}
Then we can obtain a closed solution of the problem for any functional form of $h(\cdot)$ in the Laplace-Laplace domain:
\begin{eqnarray}
\hat{\hat{\mathcal{F}}}(s;r)=\frac{1}{s+\Lambda+\lambda \left[1-\hat{h}(r)\right]+\Gamma r}\cdot\frac{\hat{\mathcal{A}}(s)}{r}.
\label{Closed_P}
\end{eqnarray}

\subsection{Pure drift}
Let us exemplify the general result in Eq.~(\ref{Closed_P}) considering the case $\lambda=0$, for which 
\begin{eqnarray}
\hat{\hat{\mathcal{F}}}(s;r)=\frac{1}{s+\Lambda+\Gamma r}\cdot\frac{\hat{\mathcal{A}}(s)}{r}.
\label{FL_l_zero}
\end{eqnarray}
Even this example could seem too trivial |when $\lambda=0$, the survival probability can be directly evaluated from Eq.~(\ref{Full_PL})| the solution to be introduced also applies to the case in which one has a jump distribution that is highly peaked around its mean, since then 
\begin{equation}
\lambda \left[1-\hat{h}(r)\right]\sim \lambda\, \EE[J]\, r, 
\end{equation}
and the corresponding results are obtained upon the replacement $\Gamma \mapsto \Gamma + \lambda\, \EE[J]$.

When $\lambda=0$, Eq.~\eqref{FL_l_zero} leads to
\begin{eqnarray}
\hat{\mathcal{F}}(s;z)&=&\frac{\hat{\mathcal{A}}(s)}{s+\Lambda}\left[1- e^{-\frac{s+\Lambda}{\Gamma}z}\right],
%\label{Full_FL}
\end{eqnarray}
and then
\begin{equation}
\hat{\mathcal{P}}_{\lint}(s;x,+)=\frac{1- e^{-\frac{s+\Lambda}{\Gamma}(\ell-x)}}{s+\Lambda\rho}\left[1+ \Lambda\rho\, \hat{\mathcal{P}}_{\lint}(s;0,+)\right]. 
%\label{Full_FL}
\end{equation}
From this
\begin{equation}
\hat{\mathcal{P}}_{\lint}(s;0,+)=\frac{1- e^{-\frac{s+\Lambda}{\Gamma}\ell}}{s+\Lambda\rho\, e^{-\frac{s+\Lambda}{\Gamma}\ell}},%\label{Full_FL}
\end{equation}
and finally
\begin{eqnarray}
\hat{\mathcal{P}}_{\lint}(s;x,+)&=&\frac{1- e^{-\frac{s+\Lambda}{\Gamma}(\ell-x)}}{s+\Lambda\rho\, e^{-\frac{s+\Lambda}{\Gamma}\ell}},\\
%\label{Full_PLP}
\hat{\mathcal{P}}_{\lint}(s;x,-)&=&\frac{1}{s+\Lambda\rho\, e^{-\frac{s+\Lambda}{\Gamma}\ell}},
%\label{Full_PLM}
\end{eqnarray}
which in turn implies
\begin{eqnarray}
\TT_{\lint}(x,+)&=&\frac{e^{\frac{\Lambda}{\Gamma}\ell}- e^{\frac{\Lambda}{\Gamma}x}}{\Lambda\rho},\\
\TT_{\lint}(x,-)&=&\frac{1}{\Lambda\rho}e^{\frac{\Lambda}{\Gamma}\ell}.
%\label{Full_TL}
\end{eqnarray}
The unconditional MFPT, starting from $x=0$ is thus
\begin{equation}
\TT_{\lint}(0)=\rho \TT_{\lint}(0,+) + (1-\rho) \TT_{\lint}(0,-)=\frac{1}{\Lambda}\left[\frac{1}{\rho} e^{\frac{\Lambda}{\Gamma}\ell}-1\right],
\label{TT_nolambda}
\end{equation}
which increases exponentially with $\ell$, $\Gamma^{-1}$, and $\Lambda$. 

Let us now assume that the value of $\Lambda$ can be somehow tuned to minimize the MFPT. Note that, for the analysis to be meaningful, one must also assume that the evolution of the system cannot be directly observed and, in particular, that one gets no notice about when the particle reaches the boundary: After a lapse of $\ell/\Gamma$, a rational agent will know with certainty that the system is not approaching the boundary, and she or he will immediately trigger the reset event. This excludes a Poissonian approach as the optimal resetting method, and leads to
\begin{equation}
\TT^{\text{rat.}}_{\lint}(0)=\frac{\ell}{\Gamma\rho}.
\label{TT_rat}
\end{equation}

\begin{figure}[htbp]
\includegraphics[scale=0.65]{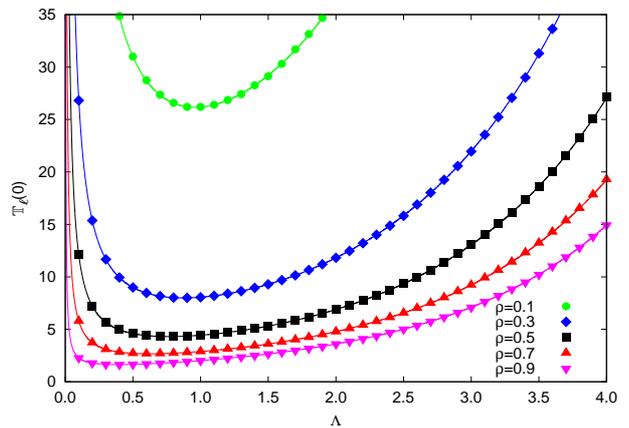}
\caption{Mean first-passage time $\TT_{\lint}(0)$ as a function of $\Lambda$ when $\lambda=0$, for different values of $\rho$. Solid lines are the direct representation of the analytical result, Eq.~(\ref{TT_nolambda}), whereas points were obtained by averaging 1\,000\,000 numerical simulations of the process. The presence of an optimal choice for $\Lambda$ becomes evident. Time variables are expressed in $\ell/\Gamma$ units.}
\label{fig:1} 
\end{figure}

\begin{figure}[htbp]
\includegraphics[scale=0.65]{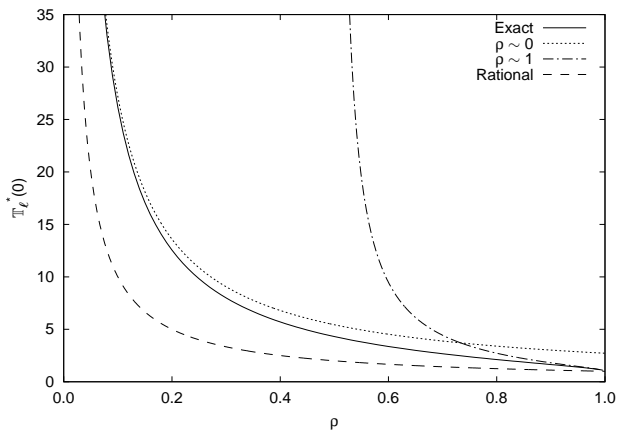}
\caption{Minimum mean exit time as a function of $\rho$ when $\lambda=0$. The solid line is the exact optimal mean exit time, obtained by solving Eq.~(\ref{TT_transcendent}); the dotted line corresponds to the approximation for small values of $\rho$,  Eq.~(\ref{rho_small}); the dotted-dashed line corresponds to the approximation for values of $\rho$ close to unity,  Eq.~(\ref{rho_large}); and the dashed line corresponds to the rational strategy, Eq.~\eqref{TT_rat}.  $\TT^*_{\lint}(0)$ is expressed in $\ell/\Gamma$ units.}
\label{fig:2} 
\end{figure}

Equation~\eqref{TT_nolambda} can be minimized by changing $\Lambda$, see Fig~\ref{fig:1}. Specifically, the optimal reset frequency $\Lambda^*$ solves the transcendental equation
\begin{equation}
e^{\frac{\Lambda^*}{\Gamma}\ell}\left(\frac{\Lambda^*}{\Gamma}\ell -1\right) +\rho =0,
\end{equation}
that is
\begin{equation}
e^{\xi_{\rho}}\left(\xi_{\rho} -1\right) +\rho =0.
\label{TT_transcendent}
\end{equation}
Some approximate solutions of Eq.~\eqref{TT_transcendent} can be obtained, see Fig~\ref{fig:2}. When $\rho \simeq 0$ one has
\begin{equation}
\Lambda^*\simeq \frac{\Gamma}{\ell}\left(1-\frac{\rho}{e}\right),
\end{equation}
and
\begin{equation}
\TT^*_{\lint}(0)\simeq \frac{\ell}{\Gamma}\cdot\frac{e}{\rho}=\TT^{\text{rat.}}_{\lint}(0)\cdot e.
\label{rho_small}
\end{equation}
When $\rho \simeq 1$ one has
\begin{equation}
\Lambda^*\simeq \frac{\Gamma}{\ell}\sqrt{2(1-\rho)},
\end{equation}
and
\begin{equation}
\TT^*_{\lint}(0)\simeq \frac{\ell}{\Gamma}\cdot\frac{1}{1-\sqrt{2(1-\rho)}}\geq  \frac{\ell}{\Gamma}.
\label{rho_large}
\end{equation}
%Noticeably peculiar is the case when $\rho=0.5\, e^{0.5}\approx 0.82436$. Under this circumstance
%\begin{equation}
%\TT^*_{\lint}(0)=\frac{1}{\Lambda^*}= 2\frac{\ell}{\Gamma},
%\end{equation}
%that is, the optimal mean first-passage time coincides with the optimal mean reset period, and duplicates the ballistic exit time: the rational optimal time when $\rho=1/2$.

\subsection{Exponential jumps without drift}
In this second example we will consider that the jumps are exponentially distributed and there is no drift, i.e., that $h(u)=\gamma e^{-\gamma u}$ and $\Gamma=0$. In this case, Eq.~\eqref{Closed_P} reads
\begin{eqnarray}
\hat{\hat{\mathcal{F}}}(s;r)&=&\frac{\gamma +r}{\left(s+\Lambda +\lambda\right) r + \gamma\left(s+\Lambda\right)}\cdot \frac{\hat{\mathcal{A}}(s)}{r}\nonumber\\
&=&\frac{\hat{\mathcal{A}}(s)}{s+\Lambda} \left[\frac{1}{r}-\frac{\lambda}{\left(s+\Lambda +\lambda\right) r + \gamma\left(s+\Lambda\right)}\right].\nonumber \\
\label{Exp_a}
\end{eqnarray}
The $r$-Laplace inversion of Eq.~(\ref{Exp_a}) is
\begin{eqnarray}
\hat{\mathcal{F}}(s;z)
&=&\frac{\hat{\mathcal{A}}(s)}{s+\Lambda} \left[1-\frac{\lambda  e^{-\alpha_s z}}{s+\Lambda +\lambda}\right],
\label{Exp_b}
\end{eqnarray}
with
\begin{equation}
\alpha_s=\frac{ \gamma\left(s+\Lambda\right)}{s+\Lambda +\lambda}.
\end{equation}
Then
\begin{equation}
\hat{\mathcal{P}}_{\lint}(s;x,+)=
\frac{1+ \Lambda\rho\, \hat{\mathcal{P}}_{\lint}(s;0,+)}{s+\Lambda\rho}  \left[1-\frac{\lambda  e^{-\alpha_s (\ell-x)}}{s+\Lambda +\lambda}\right].
\label{Exp_c}
\end{equation}
The value of $\hat{\mathcal{P}}_{\lint}(s;0,+)$ is obtained after demanding self-consistency to expression (\ref{Exp_c}) by letting $x=0$, 
\begin{equation}
\hat{\mathcal{P}}_{\lint}(s;0,+)=
\frac{s+\Lambda+\lambda\left(1-e^{-\alpha_s \ell}\right)}{s\left(s+\Lambda +\lambda\right)+\Lambda \lambda \rho\, e^{-\alpha_s \ell}}.
%\label{Exp_c}
\end{equation}
From this we get 
\begin{eqnarray}
\TT_{\lint}(0,+)&=&\frac{1}{\Lambda  \rho}\left[\frac{\Lambda + \lambda}{\lambda}e^{\alpha_0 \ell}-1\right],\\
\TT_{\lint}(0,-)&=&\frac{\Lambda + \lambda}{\Lambda  \lambda\rho}e^{\alpha_0 \ell},
%\label{Full_TL}
\end{eqnarray}
where
\begin{equation}
\alpha_0=\frac{\gamma\Lambda}{\Lambda +\lambda}.
\end{equation}
The unconditional MFPT, starting from $x=0$ is
\begin{equation}
\TT_{\lint}(0)=\frac{1}{\Lambda}\left[\frac{\Lambda + \lambda}{\lambda\rho}e^{\alpha_0 \ell}-1\right],
\label{TT_nogamma}
\end{equation}
an expression that increases exponentially with $\ell$ and $\gamma$, but tends to a fixed quantity for large values of $\Lambda/\lambda$:
\begin{equation}
\lim_{\frac{\Lambda}{\lambda}\to \infty}\TT_{\lint}(0)=\frac{1}{\lambda \rho}e^{\gamma \ell}.
\end{equation}
 
\begin{figure}[htbp]
\includegraphics[scale=0.65]{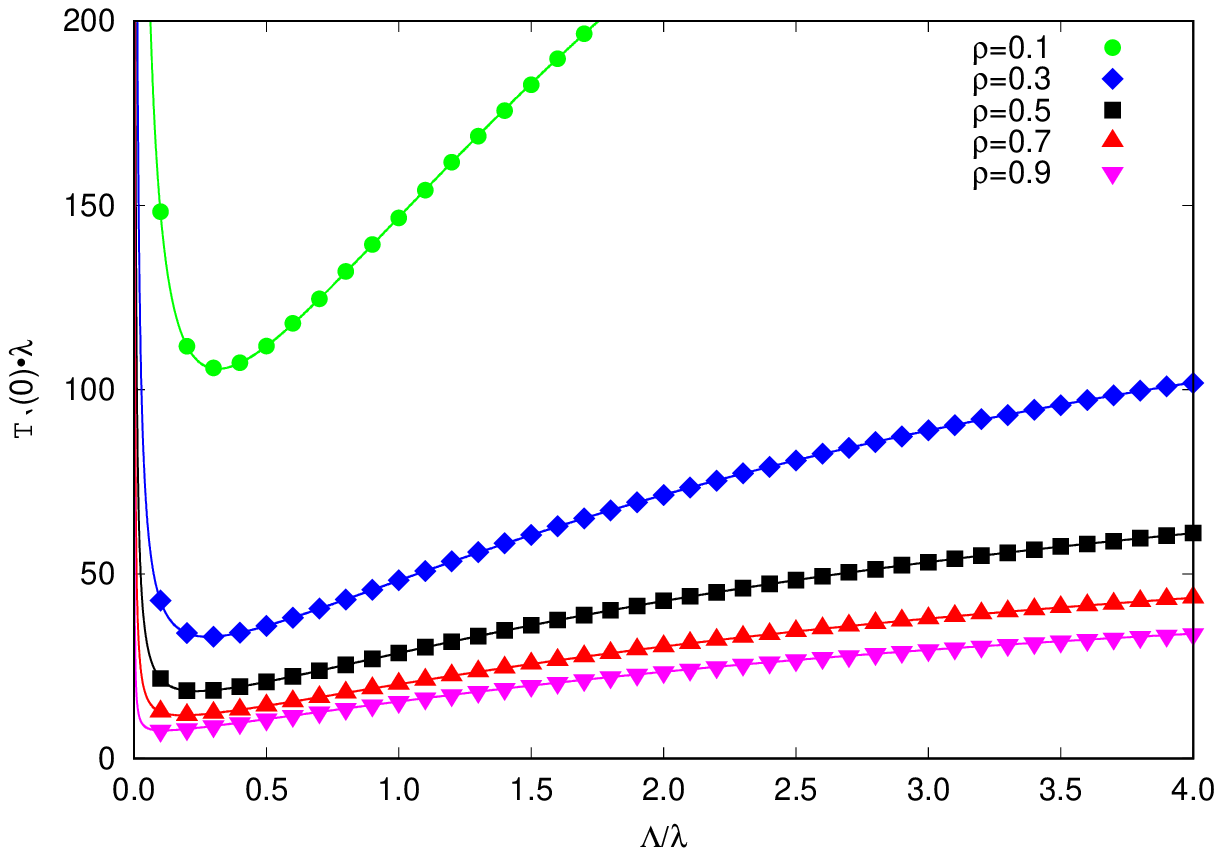}
\caption{Mean first-passage time $\TT_{\lint}(0)$ as a function of $\Lambda/\lambda$ when $\Gamma=0$ and  %the jumps are exponentially distributed, for
 $\gamma \ell =4$, for different values of $\rho$. Solid lines are the direct representation of the analytical result, Eq.~(\ref{TT_nogamma}), whereas points were obtained by averaging 1\,000\,000 numerical simulations of the process. %In this case $\gamma \ell =4>1$, then there exists an optimal value for all values of $\rho$.
}
\label{fig:3} 
\end{figure}

Again, the MFPT can be minimized with the proper choice of $\Gamma$, see Fig.~\ref{fig:3}. In fact, the equation that determines the optimal reset frequency $\Lambda^*$ is
\begin{equation}
e^{\alpha_0 \ell}\left(\alpha_0 \ell -1\right) +\rho =0,
\end{equation}
formally the same expression as before, cf. Eq.~(\ref{TT_transcendent}). Of course, the value of $\Lambda^*$ will be different,
\begin{equation}
\frac{\Lambda^*}{\lambda}=\frac{\xi_{\rho}}{\gamma \ell -\xi_{\rho}}.
\label{Lambda_opt}
\end{equation} 

Equation~\eqref{Lambda_opt} has two main consequences. The most direct one is that we must recalculate the approximate expressions that we have previously obtained. Thus, for $\rho \simeq 0$ one has
\begin{equation}
\frac{\Lambda^*}{\lambda}\simeq \frac{e-\rho}{\left(\gamma \ell-1\right) e + \rho},
\end{equation}
and
\begin{equation}
\TT^*_{\lint}(0)\simeq\frac{\gamma \ell}{\lambda}   \cdot  \frac{e}{\rho},
\label{rho_small_2}
\end{equation}
while for $\rho \simeq 1$ one has
\begin{equation}
\frac{\Lambda^*}{\lambda}\simeq  \frac{\sqrt{2(1-\rho)}}{\gamma\ell -\sqrt{2(1-\rho)}},
\end{equation}
and
\begin{equation}
\TT^*_{\lint}(0)\simeq \frac{1}{\lambda}\cdot\left(\frac{\gamma\ell}{1-\sqrt{2(1-\rho)}}+1\right).% \geq  \frac{\ell}{\Gamma}.
\label{rho_large_2}
\end{equation}
%check Fig.~\ref{fig:4}.

%\begin{figure}[htbp]
%\includegraphics[scale=0.65]{Fig4.eps}
%\caption{Optimal exit time as a function of $\rho$ when $\Gamma=0$ and  $\gamma \ell =4$. The solid line is the exact optimal mean first-passage time, obtained from Eq.~(\ref{Lambda_opt}); the dotted line corresponds to $\rho\simeq 0$,  Eq.~(\ref{rho_small_2}); and the dotted-dashed line corresponds to $\rho\simeq 1$,  Eq.~(\ref{rho_large_2}).}
%\label{fig:4} 
%\end{figure}

\begin{figure}[htbp]
\includegraphics[scale=0.65]{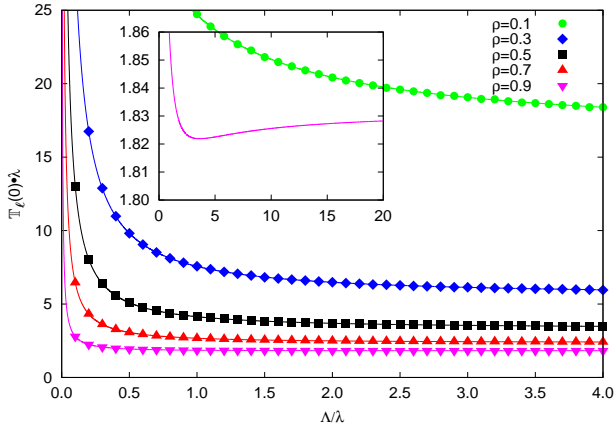}
\caption{Mean first-passage time $\TT_{\lint}(0)$ as a function of $\Lambda/\lambda$ when $\Gamma=0$ and $\gamma \ell =0.5$, for different values of $\rho$. Solid lines are the direct representation analytical results, Eq.~(\ref{TT_nogamma}), whereas points were obtained by averaging 1\,000\,000 numerical simulations of the process.}
\label{fig:5} 
\end{figure}

The second consequence of Eq.~\eqref{Lambda_opt} is more significant. Since $\xi_{\rho}\in[0,1]$ when $\rho\in[0,1]$, this means that one has an optimal value of $\Lambda$ for every $\rho$ if and only if $\gamma \ell >1$, as in Fig.~\ref{fig:3}, while for $\gamma \ell \leq 1$ there is no such optimal choice when $\xi_{\rho}\geq \gamma\ell$. In Fig.~\ref{fig:5} we illustrate this effect by setting $\gamma \ell =1/2$. With this proviso, there exist optimal choices for $\Lambda$ only when $\rho>0.5\, e^{0.5}\approx 0.82436$. This implies that with the sole exception of the line corresponding to $\rho=0.9$, see the inset in Fig. \ref{fig:5}, all the curves are monotonically decreasing for increasing values of $\Lambda/\lambda$, exhibiting no minimum: when $\xi_{\rho} \geq \gamma \ell$, the optimal strategy relies on crossing the threshold  by a single hop from the origin, and the number of such trials increases as $\Lambda/\lambda \to \infty$. 

\section{Conclusions}
\label{sec:conclusions}
In this paper we have considered the consequences of the inclusion of reset events in the dynamics of continuos-time random walks: We have reviewed some of the most outstanding contributions with connections to this topic; we have introduced a general framework to address problems of this kind, and we have presented new results corresponding to a monotonic continuous-time random walk with a constant drift, which may experience a change of orientation after every restart. We found that the system that results from this dichotomic behavior has remarkable properties

We obtained the transition probability density function for any choice of jump densities and drift speeds, and show how a stationary distribution does appear. Similarly, a formula for the mean first-passage time of the model was determined. This is a key magnitude in detection-oriented problems, and we analyze how and when this time is able to be minimized. 

Finally, we want to stress that the reported analytic expressions are found to be in good agreement with the numerical results obtained using Monte Carlo methods to average over a large number of realizations.

%\section*{Acknowledgements}
\begin{acknowledgements}
M. M. and J. V. acknowledge funding from the Spanish Agencia Estatal de Investigaci\'on and from the Fondo Europeo de Desarrollo Regional  (AEI/FEDER, UE) under Contract No. FIS2016-78904-C3-2-P. M. M. also thanks the Catalan Ag\`encia de Gesti\'o d'Ajuts Universitaris i de Recerca (AGAUR), under Contract No. 2014SGR608. A. M.-P. acknowledges support from the Spanish Ministerio de Educaci\'on, Cultura y Deporte (MECD).
\end{acknowledgements}

All authors contributed equally to this paper.


\begin{thebibliography}{99}

\bibitem{Sca07} \Journal{E. Scalas}{2006}{The application of continuous-time random walks in finance and economics}{Physica A}{362}{225}{-239} %%
   
\bibitem{Vic14} \Book{V. M\'endez, D. Campos, F. Bartumeus}{2014}{Stochastic Foundations in Movement Ecology}{Springer-Verlag}{Berlin} %
   
\bibitem{Hof13} \Journal{F. H\"ofling, T. Franosch}{2013}{Anomalous transport in the crowded world of biological cells}{Rep. Prog. Phys.}{76}{046602}{} %%

\bibitem{Lev77} \Journal{B. Levikson}{1977}{The age distribution of Markov processes}{J. Appl. Probab.}{14}{492}{-506} %

\bibitem{Pak78} \Journal{A.G. Pakes}{1978}{On the age distribution of Markov processes}{J. Appl. Probab.}{15}{65}{-77} %

\bibitem{Pak79} \Journal{A.G. Pakes}{1979}{The age of a Markov processes}{Stoc. Proc. Their Appl.}{8}{277}{-303} %%

\bibitem{Kir94} \Journal{E.G. Kyriakidis}{1994}{Stationary probabilities for a simple immigration-birth-death process under the influence of total catastrophes}{Stat. Probab. Lett.}{20}{239}{-240} %

\bibitem{Swi01} \Journal{R.J. Swift}{2001}{Transient probabilities for a simple birth-death-immigration process under the influence of total catastrophes}{Internat. J. Math. Math. Sci.}{25}{689}{-692} %%

\bibitem{Cha03} \Journal{X. Chao, Y. Zheng}{2003}{Transient analysis of an immigration birth-death process with total catastrophes}{Probab. Eng. Inform. Sci.}{17}{83}{-106} %

\bibitem{Kri00} \Journal{B. Krishna Kumar, D. Arivudainambi}{2000}{Transient solution of an M/M/1 queue with catastrophes}{Comput. Math. Appl.}{40}{1233}{-1240} %%

\bibitem{DiC03} \Journal{A. Di Crescenzo, V. Giorno, A.G. Nobile, L.M. Ricciardi}{2003}{On the M/M/1 queue with catastrophes and its continuous approximation}{Queuing Syst.}{43}{329}{-347} %%

\bibitem{LSZ93} \Journal{M. Luby, A. Sinclair, D. Zuckerman}{1993}{Optimal speedup of Las Vegas algorithms}{Inf. Process. Lett.}{47}{173}{-180} %

%\bibitem{Kau02} \Journal{Kautz H., Horvitz E., Ruan Y., Gomes C. and Selman B.}{2002}{Dynamic Restart Policies}{Proceedings of the 18th national conference on artificial intelligence}{AAAI-02}{674}{681}
\bibitem{Kau02} H. Kautz, E. Horvitz, Y. Ruan, C. Gomes, B. Selman, in \textit{Proceedings of the 18th national conference on artificial intelligence (AAAI-02), Edmonton, Alberta, Canada, 2002}, (AAAI Press, Cambridge, 2002), p. 674 %%

%\bibitem{Hua06}  \Journal{J. Huang}{2007}{The Effect of Restarts on the Efficiency of Clause Learning}{Proceedings of the 20th international joint conference on artificial intelligence}{IJCAI-07}{2318}{2323}
\bibitem{Hua06}  J. Huang, in \textit{Proceedings of the 20th international joint conference on artificial intelligence (IJCAI-07), Hyderabad, India, 2007}, edited by  R. Sangal, H. Mehta, R. K. Bagga, (Morgan Kaufmann Publishers, San Francisco, 2007), p. 2318 %%

\bibitem{Man99} \Journal{S.C. Manrubia, D.H. Zanette}{1999}{Stochastic multiplicative processes with reset events}{Phys. Rev. E}{59}{4945}{-4948} %%

\bibitem{Eva11_1} \Journal{M.R. Evans, S.N. Majumdar}{2011}{Diffusion with stochastic resetting}{Phys. Rev. Lett.}{106}{160601}{} %%

\bibitem{Eva11_2} \Journal{M.R. Evans, S.N. Majumdar}{2011}{Diffusion with optimal resetting}{J. Phys. A-Math. Theor.}{44}{435001}{} %%

\bibitem{Eva13_2} \Journal{J. Whitehouse, M.R. Evans, S.N. Majumdar}{2013}{Effect of partial absorption on diffusion with resetting}{Phys. Rev. E}{87}{022118}{} %%

\bibitem{Eva14_1} \Journal{M.R. Evans, S.N. Majumdar}{2014}{Diffusion with resetting in arbitrary spatial dimension}{J. Phys. A-Math. Theor.}{47}{285001}{} %%

\bibitem{Pal14}  \Journal{A. Pal}{2015}{Diffusion in a potential landscape with stochastic resetting}{Phys. Rev. E}{91}{012113}{} %%

\bibitem{Pal15} \Journal{A. Pal, A. Kundu, M.R. Evans}{2016}{Diffusion under time-dependent resetting}{J. Phys. A-Math. Theor.}{49}{225001}{} %%

\bibitem{Eul15} \Journal{S. Eule, J.J. Metzger }{2016}{Non-equilibrium steady states of stochastic processes with intermittent resetting}{New J. Phys}{18}{033006}{} %%

\bibitem{Tou15} \Journal{J.M. Meylahn, S. Sabhapandit, H. Touchette}{2015}{Large deviations for Markov processes with resetting}{Phys. Rev. E}{92}{062148}{} %%

\bibitem{Bha16} \Journal{U. Bhat, D. De Bacco, S. Redner}{2016}{Stochastic Search with Poisson and Deterministic Resetting}{J. Stat. Mech.-Theory Exp.}{2016}{083401}{} %%

\bibitem{Boy17} \Journal{D. Boyer, M.R. Evans M.R., S.N. Majumdar}{2017}{Long time scaling behaviour for diffusion with resetting and memory}{J. Stat. Mech.-Theory Exp.}{2017}{023208}{} %%

\bibitem{Reu16} \Journal{S. Reuveni}{2016}{Optimal stochastic restart renders fluctuations in first passage times universal}{Phys. Rev. Lett.}{116}{170601}{} %%

\bibitem{Pal17} \Journal{A. Pal, S. Reuveni}{2017}{First passage under restart}{Phys. Rev. Lett.}{118}{030603}{} %%

\bibitem{Fal16} \Journal{R. Falcao, M.R. Evans}{2017}{Interacting brownian motion with resetting}{J. Stat. Mech.-Theory Exp.}{2017}{023204}{} %%

\bibitem{Jan12} \Journal{S. Janson, Y. Peres}{2012}{Hitting times for random walks with restarts}{SIAM J. Discrete Math.}{26}{537}{-547} %%

\bibitem{Boy14} \Journal{D. Boyer, C. Solis-Salas}{2014}{Random walks with preferential relocations to places visited in the past and their application to biology}{Phys. Rev. Lett.}{112}{240601}{} %%

\bibitem{Maj15} \Journal{S.N. Majumdar, S. Sabhapandit, G. Schehr}{2015}{Random walk with random resetting to the maximum}{Phys. Rev. E}{92}{052126}{} %%

\bibitem{Miq16} \Journal{M. Montero, J. Villarroel}{2016}{Sisyphus random walk}{Phys. Rev. E}{94}{032132}{} %%

\bibitem{Kus14} \Journal{\L. Ku\'smierz, S.N. Majumdar, S. Sabhapandit, G. Schehr}{2014}{First order transition for the optimal search time of Lévy flights with resetting}{Phys. Rev. Lett.}{113}{220602}{} %%

\bibitem{Kus15} \Journal{\L. Ku\'smierz, E. Gudowska-Nowak}{2015}{Optimal first-arrival times in L\'evy flights with resetting}{Phys. Rev. E}{92}{052127}{} %%

\bibitem{Vic15} \Journal{D. Campos, V. M\'endez}{2015}{Phase transitions in optimal search times: how random walkers should combine resetting and flight scales}{Phys. Rev. E}{92}{062115}{} %%

\bibitem{Dur13} \Journal{X. Durang, M. Henkel, H. Park}{2014}{The statistical mechanics of the coagulation-diffusion process with a stochastic reset} {J. Phys. A-Math. Theor.}{47}{045002}{} %%

\bibitem{Rol16} \Journal{\'E. Rold\'an, A. Lisica, D. S\'anchez-Taltavull, S.W. Grill}{2016}{Stochastic resetting in backtrack recovery by RNA polymerases}{Phys. Rev. E}{93}{062411}{} %%

\bibitem{Fuc16} \Journal{J. Fuchs, S. Goldt, U. Seifert }{2016}{Stochastic thermodynamics of resetting}{Europhys. Lett.}{113}{60009}{} %%

\bibitem{Rot15} \Journal{T. Rotbart, S. Reuveni, M. Urbakh}{2015}{Michaelis-Menten reaction scheme as a unified approach towards the optimal restart problem}{Phys. Rev. E}{92}{060101(R)}{} %%

%\bibitem{Hus16}  K. Husain, S. Krishna, arXiv:1609.03754v1 Efficiency of a Stochastic Search with Punctual and Costly Restarts. \textit{To be published.}

\bibitem{Miq13} \Journal{M. Montero, J. Villarroel}{2013}{Monotonic continuous-time random walks with drift and stochastic reset events}{Phys. Rev. E}{87}{012116}{} %%

\bibitem{Vic16} \Journal{V. M\'endez, D. Campos}{2016}{Characterization of stationary states in random walks with stochastic resetting}{Phys. Rev. E}{93}{022106}{} %%

\end{thebibliography}
\end{document}